# Probing brain oxygenation
# with near infrared spectroscopy

**Alexander Gersten[1], Jacqueline Perle[2], Amir Raz[3] and Robert Fried[2]**

[1]Department of Physics, Ben-Gurion University of the Negev, Be'er Sheva, Israel.
 e-mail: alex.gersten@gmail.com

[2]Department of Psychology, Hunter College of the City University of New York, NY, USA

[3]McConnell Brain Imaging Centre, Montreal Neurological Institute, McGill University, Montreal, Canada

## Abstract

The fundamentals of near infrared spectroscopy (NIRS) are reviewed. This technique allows to measure the oxygenation of the brain tissue. The particular problems involved in detecting regional brain oxygenation ($rSO_2$) are discussed. The dominant chromophore (light absorber) in tissue is water. Only in the NIR light region of 650-1000 nm, the overall absorption is sufficiently low, and the NIR light can be detected across a thick layer of tissues, among them the skin, the scull and the brain. In this region, there are many absorbing light chromophores, but only three are important as far as the oxygenation is concerned. They are the hemoglobin ($HbO_2$), the deoxy-hemoglobin (Hb) and cytochrome oxidase (CtOx). In the last 20 years there was an enormous growth in the instrumentation and applications of NIRS. . The devices that were used in our experiments were : Somanetics's INVOS Brain Oximeter (IBO) and Toomim's HEG spectrophotometer. The performances of both devices were compared including their merits and drawbacks. The IBO is based on extensive efforts of an R&D group to develop a reliable device, which measures well the $rSO_2$. It is now used efficiently in operating rooms, saving human lives and expenses. Its use for research however has two drawbacks: the sampling rate is too small and the readings are limited to only two significant digits. The HEG device does not have these drawbacks, but is not developed sufficiently at this time to measure $rSO_2$. We have measured the HEG readings and compared them with the $rSO_2$ readings of the IBO. Our findings show that the HEG can be used to measure relative changes of $rSO_2$.

# Introduction

The study of the human brain made a big step forward with the introduction of noninvasive techniques, among them the near-infrared  spectroscopy (NIRS). This technique allows to measure the oxygenation of the brain tissue (Alfano et al., 1997, 1998; Chance 1998, 1998a; Delpy and Cope, 1997; Hoshi, 2003; Obrig, 2003; Owen-Reece, 1999; Rolfe, 2000; Strangman et al., 2002).

The light, with wavelengths 650-1000 nm, penetrates superficial layers of the human body, among them the skin, the scull and the brain. It is either scattered within the tissue or absorbed by absorbers present in the tissue (chromophores).

The visible light has wavelengths 400-700 nm, some individuals can see up to 760 nm. Formally, the red light extends within the wavelengths of 630-760 nm, and the near infrared light within 760-1400 nm. However, these terms are not precise, and are used differently in various studies. Here we will refer to the near infrared light as having the spectrum of 650-1000 nm, i.e. the light that penetrates superficial layers of the human body.

The dominant chromophore in tissue is water. It absorbs strongly below 300 nm and above 1000 nm. The visible part of the light spectrum, between 400 and 650 nm, is almost non-transparent due to strong absorption of hemoglobin and melanin. Only in the NIR light region of 650-1000 nm, the overall absorption is sufficiently low, and the NIR light can be detected across a thick layer of tissue.

In the following pages, we will be concerned with utilizing the NIR light to determine the oxygenation of the brain tissue. In the rather transparent NIR region, there are many absorbing light chromophores, but only three are important as far as the oxygenation is concerned. They are the hemoglobin ($HbO_2$), the deoxyhemoglobin (Hb) and cytochrome oxidase (CtOx). Hb and $HbO_2$ (which carries the oxygen) are found inside the red blood cells. CtOx is the enzyme which ends the cellular respiratory chain, and is located in the mitochondrial membrane. Quantitatively important is the difference between the absorption spectra of the oxidised and reduced forms of CtOx. The concentration of cytochrome oxidase in living tissue is usually at least an order of magnitude below that of hemoglobin (Sato et al., 1976); therefore, its contribution is often neglected.

In the last 20 years there was an enormous growth in the instrumentation and applications of NIRS. A separate section will be devoted to the instrumentation. We will compare two NIRS devices: the INVOS Brain Oximeter (IBO) of Somanetics and the HEG device of H. Toomim.

## Theoretical considerations

The Beer-Lambert Law describes the absorption of light intensity in a non-scattering medium. This law states that for an absorbing compound dissolved in a non-absorbing medium, the attenuation $A$ of an incident light is proportional to the concentration $c$ of the compound in the solution and the optical pathlength $d$:

$$A = \log_{10}[I_0/I] = a \cdot c \cdot d \ . \tag{1}$$

Above, the attenuation $A$ is measured in optical densities (OD). $I_0$ is the light intensity incident on the medium. I is the light intensity transmitted through the medium. The specific extinction coefficient a of the absorbing compound is measured in 1/micromolar per cm (or in 1/milimolar per cm equal to $1000 \cdot (1/\text{micromolar})$ per cm). The concentration of the absorbing compound in the solution c is measured in micromolar (or milimolar), and

$d$ is the distance in cm in the medium that the light covers. The product $a \cdot c$ is the absorption coefficient of the medium.

In a medium containing different absorbing compounds (except at very high concentrations), the extinction coefficient is the sum of the contributions:

$$A = \log_{10}(I_0/I) = [\ a_1 \cdot c_1 + a_2 \cdot c_2 + a_3 \cdot c_3 + ... + a_n \cdot c_{nn}]\ d\ . \tag{2}$$

For a highly scattering medium, the Beer-Lambert law, should include an additive term, $G$ that describes the scattering attenuation. It should include also a multiplying factor to account for the increased optical pathlength due to scattering. The effective optical distance, called "differential pathlength" (DP), and the multiplier, called "differential pathlength factor" (DPF) are related according to:

$$DP = DPF \cdot d. \tag{3}$$

The modified Beer-Lambert law becomes:

$$A = \log_{10}(I_0/I) = [\ a_1 \cdot c_1 + a_2 \cdot c_2 + a_3 \cdot c_3 + ... + a_n \cdot c_{nn}]\ \cdot d \cdot DPF + G\ . \tag{4}$$

In Fig. 1 the specific extinction coefficients of hemoglobin ($HbO_2$), the deoxyhemoglobin (Hb) and cytochrome oxidase (CtOx) are displayed. As was mentioned before, the concentration of cytochrome oxidase in living tissue is usually at least an order of magnitude below that of hemoglobin (Sato et al., 1976); therefore, its contribution is relatively very small. Typical value for hemoglobin concentration in brain tissue is 84 micromolars.

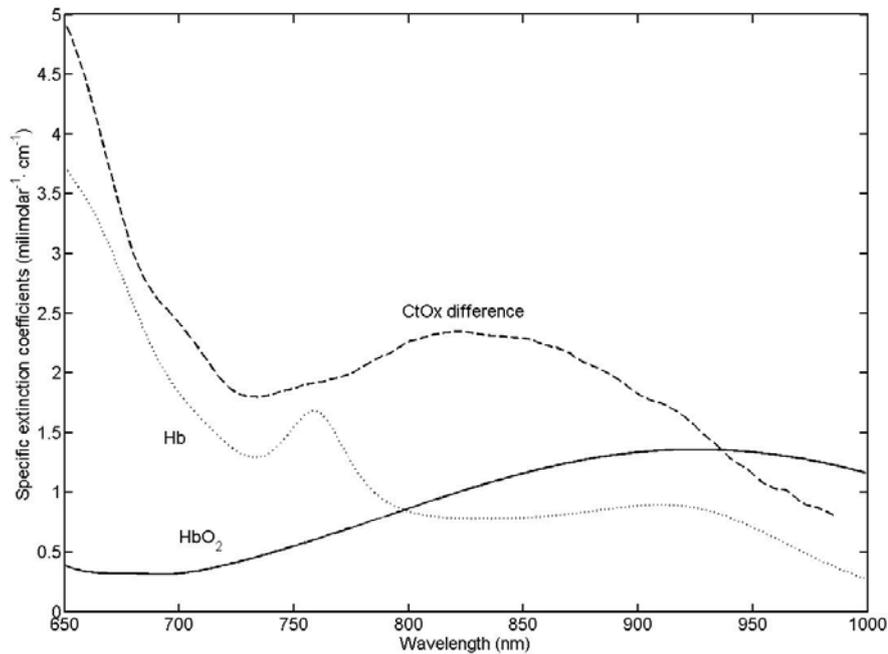

**Figure 1. The absorption spectra of HbO₂, Hb, and the difference absorption spectrum between the oxidised and reduced forms of CtOx. The data were taken from (Cope, 1991).**

It appears that DPF depends on the human age. Duncan et al. (Duncan et al., 1996) measured cranial DPF in 283 subjects whose age ranged from 1 day to 50 years. The results suggest an age dependence of DPF:

$$DPF_{780} = 5.13 + (0.07Y)^{0.81}, \tag{5}$$

where $DPF_{780}$ is the PDF at 780 nm and $Y$ is the age of the subject in years.

## Oxygen utilization

The amount of oxygen in the arterial blood depends upon the inspired oxygen and the pulmonary gas exchange. It depends on the arterial blood gas partial pressures of oxygen ($PaO_2$) and carbon dioxide ($PaCO_2$). The units of partial pressures may cause some confusion. Three types of units are in use. One unit is kiloPascal (kPa) equivalent to

7.5006 mmHg (or Torr). It can be measured also in %, when 100% corresponds to atmospheric pressure of 760 mmHg (Torr), i.e 1% corresponds to 7.6 Torr (mmHg). The arterial hemoglobin saturation ($SaO2$ ) is measured in %. The normal value is about 95%. A typical oxygen carrying capacity of the blood is 19.4 ml of $O_2$ per dl of blood with 19.1 ml $O_2$/dl carried by hemoglobin and only 0.3 ml $O_2$/dl dissolved in plasma (Code, 1991). It should be noted that the oxygen delivery to the tissues is by diffusion and the hemoglobin acts as a buffer to maintain plasma's oxygen which is extracted by the tissue.

A typical averaged value for adult cerebral blood flow (CBF) is 47.7 ml/100 ml/min (Frackowiak et al., 1980) corresponding to total oxygen delivery 9.25 ml O2/100 ml/min. (Code, 1991). Typical oxygen consumption of the adult brain is 4.2 ml O2/100 ml/min (Frackowiak et al., 1980). CBF, cerebral blood volume (CBV) and cerebral oxygen extraction (COE) are significantly greater in grey matter compared to white matter in normal human adults (Lammertsma et al., 1983; code, 1991). The CBF and the cerebral blood volume (CBV) of grey matter in normal human adults is approximately 2.5 times that of white matter, while the cerebral oxygen extractions (COE) are 0.37 and 0.41 for grey and white matter respectively (Lammertsma et al., 1983; code, 1991). Only part of the arterial oxygen which arrives in the brain is absorbed and utilized. The fraction which is utilized, known as the oxygen extraction fraction (OEF), is defined as

$$OEF=( SaO_2 - SvO_2)/SaO_2 , \qquad (6)$$

where $SaO_2$ and $SvO_2$ are the arterial and venous oxygen saturations respectively. According to Derdeyn et al (Derdeyn et al, 2002) the EOF, measured in their normal control subjects, was 0.41±0.03. Assuming $SaO_2$ equal to 0.95 the $SvO_2$ will be, using Eq. (6), equal to 0.56±0.03. In the brain tissue the absorption of the oxygenated and deoxygenated hemoglobin is mostly venous. Assuming a 75% venous contribution, the brain tissue regional oxygen saturation ($rSO_2$) in the frontal region will be about 66±3%, a mean value which is observed in experiments. With a decrease of CBF there is a bigger demand for oxygen and EOF will increase (Kissack et al, 2005). Accordingly, with an increase of CBF the OEF will decrease.

## NIRS Instrumentation

Several types of NIRS equipment, based on different methods, are commercially available. They measure the concentrations of Hb, $HbO_2$ and the total hemoglobin tHb. If the redox state of CytOx is also taken into account, then measurements with 3 wavelengths has to be done. Instruments with 2 wavelengths do not evaluate the CytOx contribution.

Three types of instruments are in use according to the used method : continuous intensity, time resolved and intensity modulated. For details, see (Delpy and Cope, 1997).

Most of the commercial instruments utilize continuous wave (CW) light. In combination with the modified Lambert-Beer law it allows to measure changes in Hb and $HbO_2$. In a biological tissue, quantification of the NIRS signal is difficult. Different methods have been proposed to improve the resolution. One of them is the spatially resolved spectroscopy (SRS), which uses CW light and a multi-distance approach. With this method the $rSO_2$ (the absolute ratio of $HbO_2$ to the total Hb content-tHb), can be evaluated (Suzuki et al., 1999).
A further distinction among the instruments can be made. The simplest are the photometers, which use single-distance and CW, light, usually with one sensor (channel).

The oximeters are more sophisticated. They use multi-distance (SRS) techniques with CW and usually two sensors (channels). For details, see (Ferrari et al., 2004; Delpy and Cope, 1997; Rolfe, 2000). Recently several groups have begun to use multi-channel CW imaging systems generating images of a larger area of the subject's head with high temporal resolution up to 10 Hz (Ferrari et al., 2004; Miura et al., 2001; Obrig and Villringer, 2003; Quaresima el al., 2001a. 2002a).

In the following we will compare two instruments: the INVOS Cerebral Oximeter of Somanetics (www.somanetics.com; Thavasothy et al., 2002), and the hemoencelograph (HEG) (Toomim and Marsh, 1999; Toomim et al., 2004).

## The INVOS Oximeter

The Somanetics INVOS Cerebral Oximeter (ICO) system measures regional hemoglobin oxygen saturation ($rSO_2$) of the brain in the area underlying the sensor and uses two wavelengths, 730 and 810 nm. The sensor, ( "SomaSensor"), is applied to the forehead with an integrated medical-grade adhesive. Two sensors can be placed in the forehead near Fp1 and Fp2. The spatially resolved spectroscopy (SRS) method is applied by using in the sensor two source-detector distances: a "near" (shallow), 3 cm from the source and a "far" (deep), 4 cm from the source. Both sample almost equally the shallow layers in the tissue volumes directly under the light sources and detectors in the sensor, but the distant "far" penetrates deeper into the brain. Using the SRS method, subtraction of the near signal from the far should leave a signal originating predominantly in the brain cortex. Practically, the constant term G in the modified Beer-Lambert law Eq. (4) is removed. The measurement takes place in real time, providing an immediate indication of a change in the critical balance of oxygen delivery and oxygen consumption.

According to the producers: "Using the model at a 4 cm source-detector spacing and no signal subtraction, the overlying tissue and skull contribute, on average, about 45 percent of the signal while 55% is cerebral in origin. Subtracting the data from the 3 cm spacing (as the Oximeter does) reduces this extracerebral contribution to less than 15 percent. While the potential exists to develop an instrument that will reduce the extracerebral contribution to zero, subject-dependent variations in anatomy and physiology will likely cause variations of ±10%. While the extracerebral contribution is not zero, the noninvasive Somanetics INVOS Cerebral Oximeter provides a "predominately cerebral" measurement where over 85 percent of the signal, on average, is exclusively from the brain" (www.somanetics.com).

The INVOS Cerebral Oximeter is an important tool in surgery rooms, saving lives and expenses. The producers explain: " Declining cerebral oximeter values occur frequently in cardiac surgery and reflect the changing haemodynamic profile of the balance between brain oxygen delivery and consumption. Since low rSO2 values correlate with adverse neurological and other outcomes, continuous assessment is a valuable patient management tool. Declining or low cerebral oximeter values are corrected with simple interventions". In (www.somanetics.com) the reader will find an extensive bibliography of medical papers supporting the above claims.

**The HEG**

The hemoencephalograph (HEG) is a single-distance CW spectrophotometer, which uses NIR light with two wavelengths, 660 and 850 nm. The light source consist of closely spaced emitting diodes (LED optodes). The source and an optode light receiver are mounted on a headband. The distance between the source and receiver is 3 cm. The HEG measures the ratio of the intensity of the 660 nm light to the intensity of the 850 nm light.

Obviously, from a single datum one cannot solve the unknowns of Eq. (4). Therefore the HEG is not intended to measure $rSO_2$. Nevertheless it is an important tool in the biofeedback research.

Hershel Toomim, the inventor of HEG has noticed that he can influence the outcome by looking at the results. Since then many people were able to increase the HEG readings via such a biofeedback. The HEG became an important tool for training local brain oxygenation.

The HEG is a very sensitive device. The distance between the source and receiver is the same as the distance of the shallow detector of the Somanetics INVOS Cerebral Oximeter. Therefore the INVOS Oximeter covers larger brain tissue and is more stable and less influenced by biofeedback.

What is the HEG measuring? According to the producers: "The HEG ratio is the basis of blood flow training. A normalized basis for HEG was established using measurements at Fp1 of 154 adult attendees at professional society meetings. A normalized reference value of 100 (SD=20) was thus established and served to calibrate all further spectrophotometers". The calibration was achieved by multiplying the intensities ratio by

200. We have shown (Gersten et. al. 2006) that one can relate the readings of the HEG to rSO$_2$ and even calibrate it separately for each individual.

Using as the basis Eq. (4) we found that the ratio of the intensities should have the form

$$\text{HEG readings} = 200 \cdot I_{660}/I_{850} = a \cdot \exp(b \cdot x), \quad x \equiv rSO_2, \tag{7}$$

Where $I_{660}$, $I_{850}$ are the intensities in the receiver of the 660 and 850 nm lights respectively, $a$ and $b$ are constants which should be dependent on each individual.

We have done measurements with 19 subjects. During 5-minute intervals measurements were taken on the forehead at Fp1 with the Somanetics INVOS Cerebral Oximeter and consecutively with HEG. The results are displayed in Fig. 2. The parameters of the best fit are

$$a = 32.08 \pm 0.11 \text{ (HEG readings)}, \quad b = 0.01581 \pm 0.00151 \text{ (1/\%)}.$$

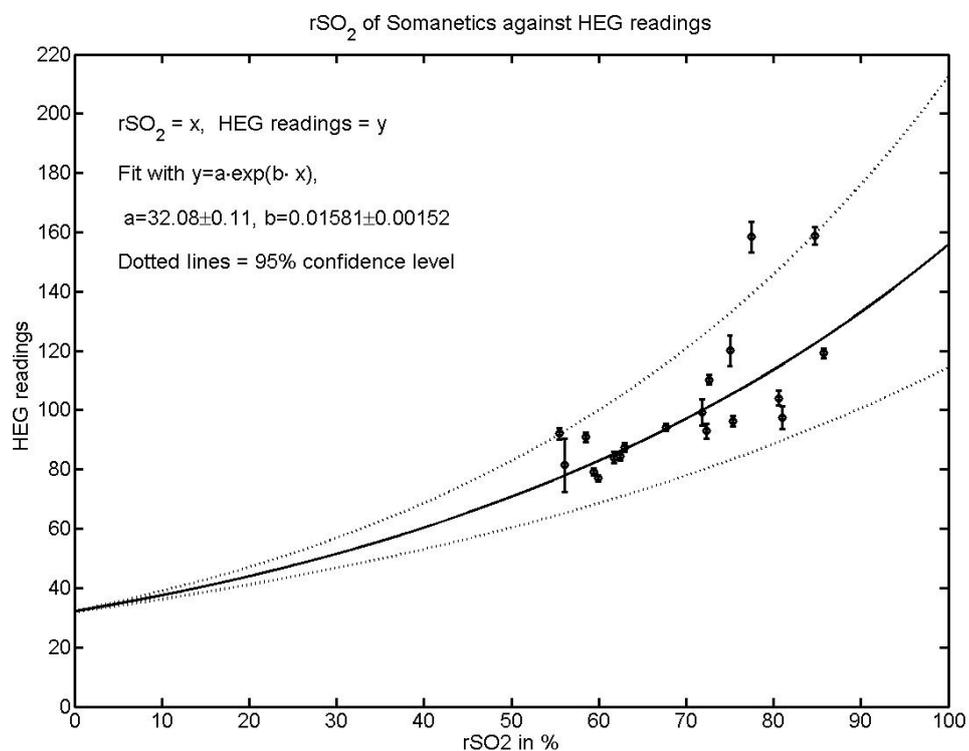

**Figure 2.** Best fit to the rSO$_2$ dependence of the HEG readings. The rSO$_2$ data were obtained with Somanetics INVOS Cerebral Oximeter.

Although the individual data in Fig. 2 are quite dispersed the error of $a$ is relatively very small and can be neglected. It means that we have practically only one constant $b$ to be determined individually. Eq.(7) can now be replaced by

HEG readings = $200 \cdot I_{660}/I_{850} = 32.08 \cdot \exp(b \cdot x)$,     $x \equiv rSO_2$.     (8)

The value of $rSO_2$ in terms of the HEG readings becomes

$$rSO_2 = [\log( \text{(HEG readings)}/ 32.08)]/b,     (9)$$

where the constant b can be calibrated for each individual using a device, which measures $rSO_2$. Other important conclusion from Eq. (9) is that relative changes of $rSO_2$ are independent on the constant $b$. Therefore, the HEG can be used to measure relative changes of $rSO_2$. Explicitly we can write

$x_1/x_2 = \log(y_1/32.08)/\log(y_2/32.08)$,  $x \equiv rSO_2$,  $y \equiv$ HEG readings.     (10)

## Other measurements and comparison of performance.

In this section, we will describe experiments done separately with the Somanetics INVOS Cerebral Oximeter and HEG (Gersten et al., 2006, 2006a, 2006b, 2007). In some experiments we could not use the Somanetics INVOS Cerebral Oximeter because of its slow sampling (sampling rate 1/12 Hz). The sampling rate of the HEG was 1 Hz (with a potential to increase it up to 122 Hz).

We have conducted experiments in which we measured end tidal $CO_2$ (with a capnometer) and brain oxygenation during breathing exercises. The measurements were done with a capnometer of Better Physiology, Ltd. The capnometer measures the $CO_2$ concentration of the expired air. During the inspiration or breath holding the capnometer indications were zero. The capnometer enabled us to follow the breathing periodicity.

The periodicity of normal breath is approximately ¼ Hz. In order to effectively use the Somanetics INVOS Cerebral Oximeter we had to study breathing exercises with very small periodicity, much smaller than 1/12 Hz. We could achieve this only because all participants were yoga practitioners. They were doing periodic breathing exercises with frequency only 3 times smaller than the 1/12 Hz. The results led us to conclude that there are oxygenation periodic waveforms in the brain with the same frequency as the breathing periodicity (Gersten et al., 2006). An example is given in Fig. 3.

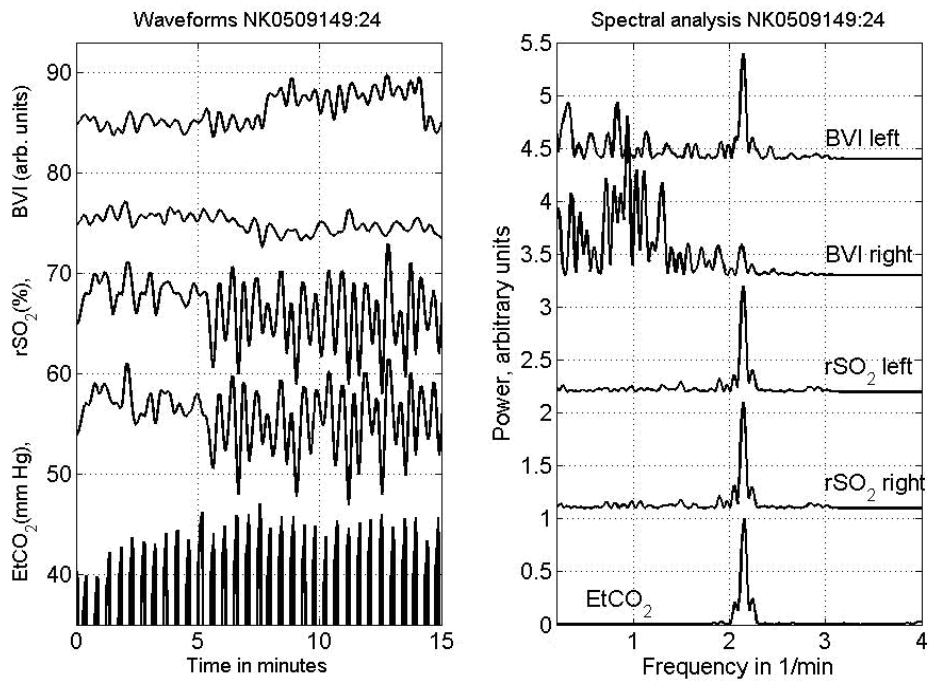

**Figure 3. On the left hand side are given from the bottom to top: the readings of the capnometer in mm Hg, $rSO_2$ from the right sensor (subtracted with 8%), $rSO_2$ from the left sensor, the difference of BVI from the right sensor, and the difference of BVI from the left sensor (increased by 10). On the right hand side the corresponding spectral analyses of the waveforms are given.**

Experiments conducted with the HEG on untrained students (Gersten et al., 2006b) revealed the oxygenation periodic waveforms with much better clarity. They did perform milder exercises with frequency about 1/12 Hz, much smaller than the sampling rate 1 Hz of the HEG. An example is given in Fig. 4.

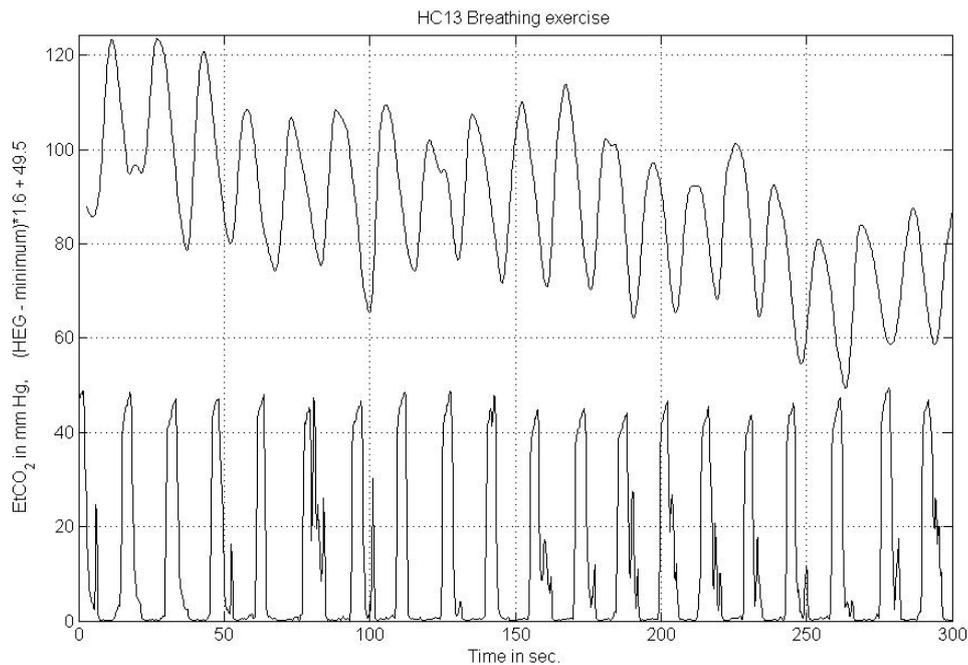

**Figure 4. The oxygenation waveform (the upper curve) obtained with the HEG. The lower curve is the $CO_2$ from the capnometer** .

## Conclusions

The near infrared spectroscopy (NIRS) is a powerful non-invasive method that probes brain and muscle oxygenation. The subject is of great interest to researchers and the medical profession, resulting in many clinical, neuroscience and physiology of exercise studies and publications.

NIRS is a noninvasive and easy to handle method, and  will provide a new direction for functional mapping. It will compete with other neuroimaging techniques, like MRI, and PET (Hoshi, 2003; Obrig and Villringer, 2003).

Two NIRS devices were used in our experiments: Somanetics's INVOS Brain Oximeter (IBO) and Toomim's HEG spectrophotometer.

The IBO is based on extensive research to develop a reliable device, which measures the $rSO_2$ well. It is used efficiently in operating rooms, saving human lives and expenses. It was also useful in many medical studies. For research it has two drawbacks: (1) its sampling rate is too small and (2) the readings, limited to two significant digits, do not allow the study of small changes.

The HEG device, which is seemingly not comparable to IBO in technology as well as in price (one order of magnitude lower) does not have the IBO's drawbacks. The HEG was not sufficiently developed to measure $rSO_2$. We have calibrated the HEG readings with the $rSO_2$ IBO's readings and found that the HEG could be used to measure relative $rSO_2$ changes. That enabled us to use the HEG effectively in research studies.

The IBO penetrates 3 cm into the brain while the HEG's is about 1.5 cm. IBO's results are more stable because it is involved with a larger brain volume. The HEG is more sensitive in the range of normal and above normal brain oxygenation, therefore it is more useful for biofeedback research.
Our measurements were done on the forehead. There is a need to continue this research into other regions of the brain.

The two devices complemented each other and allowed us to gain new insight into the brain function (for example, the effect of increasing $PaCO_2$ and the formation of oxygenation waveforms). Our expectations are that expansion of this work will lead to new understandings of brain functions.